\documentclass[11pt,a4paper]{article}
\usepackage{amsfonts}
\usepackage{amsmath}

\input{tcilatex}
\def\beq{\begin{equation}}
\def\eeq{\end{equation}}

\def\dg{\dagger}

\def\bt{\begin{tabular}}
\def\et{\end{tabular}}

\def\lam{\lambda}

\def\eps{\epsilon}

\def\rd{{\rm d}}   \def\ri{{\rm i}} \def\re{{\rm e}}

\begin{document}

\author{O. Cherbal$^{(1)}$, M. Drir$^{(1)}$, M. Maamache $^{(2)}$ and D.A.
Trifonov$^{(3)}$ \\
$^{(1)}${\normalsize \thinspace Faculty of Physics, Theoretical Physics
Laboratory, University of \ \ }\\
{\normalsize Bab-Ezzouar, USTHB, B.P. 32, El Alia, Algiers 16111, Algeria}\\
$^{(2)}$\,{\normalsize Laboratoire de Physique Quantique et Syst\`{e}mes
Dynamiques, }\\
{\normalsize Department of Physics, Setif University, Setif 19000, Algeria.}%
\\
$^{(3)}$\,{\normalsize Institute of Nuclear Research, 72 Tzarigradsko
chauss\'ee,} \\
{\normalsize 1784 Sofia, Bulgaria.}}
\title{Invariants and coherent states for a nonstationary fermionic forced
oscillator}
\maketitle

\begin{abstract}
Invariant creation and annihilation operators and related Fock states and
coherent states are built up for the system of nonstationary fermionic
forced oscillator.

\end{abstract}


\section{Introduction}

\ \ \ \ The time evolution of coherent states (CS) has attracted a great
deal of attention since the introduction of Glauber's CS of the harmonic
oscillator {\normalsize \cite{Glauber63}}. Of
particular interest has been the determination of the Hamiltonian operator
for which an initial coherent state remains coherent under time evolution.
It is established that this Hamiltonian has the form of the nonstationary
bosonic forced oscillator Hamiltonian {\normalsize \cite{Glauber66, Mehta66,
 Stoler75, Kano76}:  \ \ \ \ \ \ \ \ \ \ } \
\begin{equation}
H_{\mathrm{cs}}=\omega (t)a^{\dagger }a+f(t)a^{\dagger }+f^{\ast }(t)a+\beta
(t),  \label{Hcs}
\end{equation}%
where $\omega (t)$ and $\beta (t)$ are arbitrary real functions of time $t$,
and $f(t)$ is arbitrary complex function. 
$H_{\mathrm{cs}}$ is a particular case of the nonstationary forced oscillator
Hamiltonian for which the exact time evolution of CS has been obtained in
\cite{MMT70,Holz70} by first constructing boson ladder operator dynamical %
invariants according to the Lewis and Riesenfeld scheme of time
dependent invariants \cite{Lewis69}.

Our purpose in the present article is to study the dynamical invariants and
time evolution of CS for the {\it fermionic} forced oscillator (FFO), which %
in fact is the general (one mode) Hamiltonian.

The organization of the article is as follows. In Sec. 2 we construct fermionic %
ladder operator dynamical invariants and the corresponding Lewis-Riesenfeld
Hermitian invariant {\normalsize \cite{Lewis69}}, following the scheme related %
to the boson system {\normalsize \cite{MMT70}}. Using these invariants, we
construct in Sec. 3 fermionic CS and Fock states of FFO system as eigenstates %
of the constructed invariant fermionic annihilation operator $B(t)$ and %
$B^\dagger(t)B(t)$ correspondingly. These CS can
represent (under appropriate initial conditions) the exact time-evolution of
initial canonical fermionic CS. Finally the relation of the invariant ladder %
operators method {\normalsize \cite{MMT70, Holz70}} to the Lewis-Riesenfeld method
{\normalsize \cite{Lewis69}} is briefly described on the example of FFO. The
paper ends with concluding remarks.\

\medskip

\section{FFO and invariant ladder operators}

We consider the single nonstationary fermionic forced oscillator (FFO)
described by the following Hamiltonian,%
\begin{equation}
H_{\!f}=\omega (t)b^{\dagger }b+f(t)b^{\dagger }+f^{\ast }(t)b+g(t),
\label{Hf}
\end{equation}%
where $\omega (t)$ and $g(t)$ are arbitrary real functions of time, $f(t)$
is arbitrary complex function, $b$ and $b^{\dagger }$ are fermion
annihilation and creation operators respectively, which obey to the fermion
algebra:
\begin{equation}
\{b,b^{\dagger }\} = 1,\text{ \ }%
b^{2}=b^{\dagger }{}^{2}=0,  \label{013}
\end{equation}%
where $\{b,b^{\dagger }\}\equiv bb^{\dagger }+b^{\dagger }b$. %
Due to the nilpotency of the fermionic operators $b$, $b^\dagger$
the operator $H_{\!f}$ represents the most general (one mode) fermionic %
Hamiltonian.

The Hilbert space $\mathcal{H}$ of the single-fermion system is spanned by
the two eigenstates $\left\{ \left\vert 0\right\rangle ,\left\vert
1\right\rangle \right\} $ of  number operator $b^{\dagger }b$:\, %
$b^{\dagger }b\left\vert n\right\rangle =n\left\vert n\right\rangle ,\text{ \ }n=0,1$.
The operators $b$ and $b^{\dagger }$\ allow transitions between number states,
\begin{equation}
b\left\vert 0\right\rangle =0,\text{\ }b\left\vert 1\right\rangle
=\left\vert 0\right\rangle \,,\text{ \ }b^{\dagger }\left\vert
1\right\rangle =0,\text{ }b^{\dagger }|0\rangle =|1\rangle .\,
\end{equation}%
The form of the Hamiltonian ({\normalsize \ref{Hf}}) is a Hermitian linear
combination of $b$, $b^{\dagger }$ and $N = b^{\dagger }b$. The fermion number
operator $N$ obey the relation $N^{2}=N$ and the three operators $b$, $%
b^{\dagger }$and $N$ close under commutation the algebra:
\begin{equation}
\left[ b,N\right] =b,\text{ }\left[ b^{\dagger },N\right] =-b^{\dagger },%
\text{ }\left[ b,b^{\dagger }\right] =1-2N,\text{ \ \ \ }
\end{equation}%
Let us note that linear combinations of $b^{\dagger }$, $b$ and $N$ produce
the half-spin operators $J_{i}$,
\begin{equation}
J_{1}=\tfrac{1}{2}(b^{\dagger }+b),\quad J_{2}=\tfrac{1}{2\ri}(b^{\dagger
}-b),\quad J_{3}=b^{\dagger }b-\tfrac{1}{2},  \label{30a}
\end{equation}%
closing the \textit{su}(2) algebra: $\left[ J_{k},J_{l}\right] =i\epsilon
_{klm}J_{m}.$

It is convenient to use raising and lowering operators $J_{\pm }=J_{1}\pm
iJ_{2}$ which satisfy the following commutation relation: $\left[ J_{+},J_{-}%
\right] =2J_{3},\text{ }\left[ J_{3},J_{\pm }\right] =\pm J_{\pm }\text{\ }$%
, where $J_{+}=b^{\dagger },$ $J_{-}=b.$ So that in terms of these half spin
operators the Hamiltonian ({\normalsize \ref{Hf}}) takes the form
\begin{equation}
H_{\!f}=\omega (t)J_{3}+f(t)J_{+}+f^{\ast }(t)J_{-}+g(t)+\tfrac{\omega (t)}{2%
}.  \label{Hf 2}
\end{equation}%

Our task is the construction of the time-dependent invariants for the system
{\normalsize (\ref{Hf})}, {\normalsize (\ref{Hf 2})}. The defining equation
of the invariant operator $B(t)$ for a quantum system with Hamiltonian $H(t)$
is%
\begin{equation}
\frac{\partial }{\partial t}B(t)-i[B(t),H]=0  \label{19}
\end{equation}
Formal solutions to Eq. {\normalsize (\ref{19})} are operators $%
B(t)=U(t)B(0)U^{\dagger }(t)$, where $U(t)$ is the evolution operator of the
system, $U =T\exp [-\ri\tint_{0}^{t}H(t^{^{\prime }})\rd t^{^{\prime }}]$. 
In our case of FFO {\normalsize (\ref{Hf})}, {\normalsize (\ref{Hf 2})} we look %
for the non-Hermitian invariants $B(t)$, $B^\dagger(t)$ of the form of linear %
combination of the $SU(2)$ generators {\normalsize (\ref{30a})}, \
\begin{equation}
\begin{aligned} B = \nu_{-}(t)J_{-} + \nu_{+}(t) J_{+} + \nu_3 (t) J_{3}
,\\ B^\dg = \nu_{-}^* (t)J_{+} + \nu_{+}^*(t) J_{-} + \nu_3^* (t) J_{3},
\end{aligned}  \label{B}
\end{equation}%
where $\nu _{\pm }(t),\,\,\nu _{3}(t)$ may be complex functions of the time.
Hermitian invariants then can be easily built up as Hermitian combinations
of $B$ and $B^{\dagger }$. In particular if $B$ is a non-Hermitian invariant
the operator $I=B^{\dagger }B- 1/2$ 
is a Hermitian invariant, the fermion analog of the Lewis-Riesenfeld quadratic
invariant {\normalsize \cite{Lewis69}}.

Let us note at this point that we look for FFO invariants as elements of the
same algebra $su(2)$ to which the Hamiltonian belongs. Similar is the approach
used in \cite{Chumakov86} in construction of invariants for %
the nonstationary singular oscillator, where the related algebra is $su(1,1)$. %
This is to be compared with the case of nonsingular oscillator, for which invariant %
ladder operators have been built up as elements of the Heisenberg-Weyl algebra %
$h_w$ (i.e. as linear combinations of coordinate and momentum operators $x$ and %
$p$ \cite{MMT70,Holz70}), while the related nonstationary Hamiltonian belongs to 
$su(1,1)$. %
For {\it forced} boson oscillator the Hamiltonian belongs to the large algebra of 
semi-direct sum $su(1,1)\dot{+}h_w$ but the ladder operator invariants %
 are again elements of the invariant subalgebra $h_w$  %
\cite{Prants86,Dattoli86,Trif93}.

To proceed with construction of FFO invariants we substitute {\normalsize (\ref{B}) and 
(\ref{Hf 2}) } into {\normalsize (\ref{19})}, and find the following system of
differential equations for the parameter functions $\nu_\pm,\,\nu_3$: \
\begin{eqnarray}
\dot{\nu}_{3} &=&2\ri(\nu _{+}f^{\ast }-\nu _{-}f),  \label{(a)} \\
\dot{\nu}_{+} &=&\ri(\nu _{3}f-\nu _{+}\omega ),  \label{(b)} \\
\dot{\nu}_{-} &=&\ri(\nu _{-}\omega -\nu _{3}f^{\ast }).  \label{(c)}
\end{eqnarray}%
Solutions to the above linear system of first order equations are
uniquely determined by the initial conditions $\nu _{\pm }(0)=\nu _{0,\pm }$%
, $\nu _{3}(0)=\nu _{0,3}$. If we want the invariants $B(t)$ and $B^{\dagger
}(t)$ be again fermion ladder operators, i.e. to obey the conditions
\begin{equation}
B^{2}=0, \quad \{B,B^{\dagger}=1,  \label{constr}
\end{equation}%
we have to take $\nu _{0,\pm }$ and $\nu _{0,3}$ satisfying
\begin{equation}
 \nu _{0,3}^{2} = -4\nu _{0,+}\nu _{0,-},\quad
 |\nu _{0,-}|+|\nu _{0,+}| = 1\,. \label{0}
\end{equation}%
Indeed, for $B^{2}(t)$ and $\{B(t),B^{\dagger }(t)\}$ we find
\begin{equation}
\begin{tabular}{l}
$\displaystyle B^{2}=\nu _{+}\nu _{-}+\tfrac{1}{4}\nu _{3}^{2}\equiv
\lambda _{1}$, \\[2mm]
$\displaystyle\{B,B^{\dagger}\}=|\nu _{-}|^{2}+|\nu _{+}|^{2}+\tfrac{1%
}{2}|\nu _{3}|^{2}\equiv \lambda _{2}$.%
\end{tabular}
\label{lam_i}
\end{equation}%
The quantities $\lambda _{1}(\nu _{\pm },\nu _{3})$, $\lambda _{2}(\nu _{\pm
},\nu _{3})$ turned out to be two different '\textit{constants of motion}'
for the system {\normalsize (\ref{(a)})}-{\normalsize (\ref{(c)})}, their
time derivatives being vanishing:
\begin{equation}
\begin{tabular}{l}
$\displaystyle 
\frac{\rd}{\rd t}\lambda _{1} = \frac{\rd}{\rd t}\left( \nu _{+}\nu _{-}+\tfrac{1%
}{4}\nu _{3}^{2}\right) =0$, \\[3mm]
$\displaystyle
\frac{\rd}{\rd t}\lambda _{2} = \frac{\rd}{\rd t}\left( |\nu _{-}|^{2}+|\nu
_{+}|^{2}+\tfrac{1}{2}|\nu _{3}|^{2}\right) =0$.%
\end{tabular}
\label{dotlam_i}
\end{equation}%
Therefore we can fix the values of these constants as $\lambda _{1}=0$, $%
\lambda _{2}=1$, i.e.
\begin{equation}
\begin{tabular}{l}
$ \displaystyle \nu _{+}\nu _{-}+\tfrac{1}{4}\nu _{3}^{2}=0,$\\[3mm]
$\displaystyle |\nu _{-}|^{2}+|\nu_{+}|^{2}+\tfrac{1}{2}|\nu _{3}|^{2}=1$, 
\end{tabular}
\label{lam 0}
\end{equation}%
and satisfy the conditions {\normalsize (\ref{constr})}. If furthermore the initial
conditions are taken as
\begin{equation}
\nu _{-}(0)=1,\,\,\,\nu _{+}(0)=0=\nu _{3}(0),  \label{nu_i0}
\end{equation}%
then $B(0)=b$. Later on we work with these fermionic ladder operator invariants, %
i.e. we consider conditions (\ref{lam 0}) satisfied.

Let us now recall that in the case boson nonstationary oscillator the ladder %
operator invariants, constructed first in \cite{MMT70, Holz70} (see also %
\cite{Chumakov86, Prants86}), are expressed in terms of one only parameter function %
$\eps(t)$, which obeys a simple second order equation, namely that of the %
classical oscillator with varying frequency. It turned out that this can be done %
in the case of  fermionic oscillator as well.
In this aim we first  express all the three parameter functions $\nu _{\pm }(t)$,%
 $\nu _{3}(t)$ in terms of one of them, which has to obey a second order %
differential equation.
Let for example, express $\nu _{3}(t)$ and $\nu _{-}(t)$ in terms of $\nu _{+}(t)$%
 and its derivatives.  We have %
\begin{equation} \label{nu3}
\nu _{3} =  - \tfrac{\ri}{f}(\dot{\nu}_{+} + \ri \nu _{+}\omega ),
\end{equation}%
\begin{equation}
\nu _{-}=\tfrac{1}{2f^{2}}\left[ \ddot{\nu}_{+}+\left( \ri \omega -\tfrac{\dot{f%
}}{f}\right) \dot{\nu}_{+}+\left( 2ff^{\ast }+\ri\dot{\omega}-\ri \frac{\omega }{f%
}\dot{f}\right) \nu _{+}\right] .  \label{nu-}
\end{equation}%
Substituting these expressions into the expression of $\lam_1$ in terms of $\nu_\pm,
\,\nu_3$ and taking into account that $\lam_1$ is fixed to $0$ we find that $\nu_+$ %
should satisfy the following second order equation,

\begin{equation}
2\nu _{+}\ddot{\nu}_{+}-\dot{\nu}_{+}^{2} - 2\nu_{+}\dot{\nu}_{+}\tfrac{\dot{f}}{f} +
4\nu _{+}^{2}\left(
|f|^{2}+\tfrac{\omega ^{2}}{4}+ \ri\tfrac{\dot{\omega}}{2}-\ri \tfrac{\omega %
\dot{f }}{2f}\right)  =  0 .  \label{lam}
\end{equation}%
 Using this, and supposing that $\nu_{+}\neq 0$, we obtain for $\nu _{-}$ a more %
compact expression in terms of $\nu _{+}$ and $\dot{\nu}_{+}$, %
\begin{equation}
\nu _{-}= -\nu _{3}^{2}/4\nu _{+},  \label{nu- 2}
\end{equation}%
where $\nu_3$ is given again by eq. (\ref{nu3}).

Thus the operators $B(t),\,B^{\dagger}(t)$, eq. (\ref{B}), are fermionic %
ladder operator invariants for the forced oscillator (\ref{Hf}), (\ref{Hf 2})%
if  $\nu _{3}$ and $\nu _{-}$ are given by eqs. (\ref{nu3}) and (\ref{nu- 2}),%
and $\nu _{+}(t)$ is a nonvanishing solution of the second order equation (\ref{lam}).
%

Next we try to linearize the auxiliary eq. (\ref{lam}). In this purpose we put
\begin{equation}
\nu _{+}(t) = \tfrac{1}{2}{\eps'} ^{2}(t)  \label{nu+ eps'}
\end{equation}%
and obtain that  $\eps' (t)$ satisfies the linear equation
\begin{equation} \label{eps' eq}
\ddot{\eps}' - \tfrac{\dot{f}}{f}\dot{\eps}'+\Omega' (t)\eps' =0,
\end{equation}%
where
\begin{equation}
\Omega' (t)=|f(t)|^{2}+\tfrac{1}{4}\omega ^{2}(t)+\tfrac{\ri}{2}\dot{\omega}-%
\tfrac{\ri}{2}\omega \tfrac{\dot{f}}{f}.
\end{equation}

In terms of $\eps'$ the formulas (\ref{nu-}) and (\ref{nu3} for $\nu_-,\, \nu_3$ read
\begin{equation}
\begin{tabular}{l}
$\displaystyle\nu _{-}=-\frac{1}{2\eps'}\nu_3^{2}$, \\[2mm]
$\displaystyle\nu _{3}=\frac{1}{f}\left( \tfrac{\omega }{2}{\eps'}
^{2}-\ri\eps' \dot{\eps}'\right) $.%
\end{tabular}%
\end{equation}%
The term in (\ref{eps' eq}) proportional to the first derivative can be eliminated %
by the substitution %
\begin{equation}\label{eps}
\eps'  = \eps \exp \left( \tfrac{1}{2}\tint_{0}^{t}\rd\tau \,\dot{f}%
(\tau )/f(\tau )\right) .
\end{equation}%
This leads to the desired simple equation for $\eps$,
\begin{equation}\label{eps eq}
\ddot{\epsilon}+\Omega (t)\epsilon = 0,
\end{equation}%
where $ \Omega(t)= \Omega^{\prime }(t)  + \ddot{f}/2f - 3\dot{f}\,^{2}/4f^{2}$. %
Equation (\ref{eps eq}) is of the same type, as the auxiliary equation used in the %
case of nonstationary boson oscillator \cite{MMT70, Holz70}. Here the %
'squared frequency' $\Omega$ however is complex and depends in a different manner %
on the corresponding Hamiltonian parameters. And the solutions are subject to %
different constraints, stemming from the different commutation relations: 
in terms of our $\eps$, eq. (\ref{eps eq}), the constraint $\lam_2=1$ reads %
( $\eps'$ is related to $\eps$ according to (\ref{eps})),
\begin{equation}
\tfrac{|\eps' |^{4}}{4}\left( 1+\tfrac{2}{|f|^{2}}\left\vert
\tfrac{\omega }{2}\eps' -\ri\dot{\eps}'\right\vert ^{2}+\tfrac{1}{|f|^{4}%
}\left\vert \tfrac{\omega }{2}\eps' -\ri\dot{\eps}'\right\vert^{4}\right) =1 ,
\end{equation}%
 while in  the boson case the constraint is  ${\rm Im}\left({\eps}^*\dot{\eps}\right) = 1$ 
\cite{MMT70, Holz70}.

To finalize this section let note that in the particular case of the %
\textit{free} fermion oscillator, $f(t)\equiv 0$, the explicit solutions of the problem can %
 be easily found in the form %
\begin{equation}
\begin{tabular}{l}
$\displaystyle 
\nu _{\pm }(t)=\nu _{0,\pm }\re^{\pm \ri\tint^{t}\omega (\tau )\rd\tau },$\\
$ \displaystyle 
\nu_{3}=\nu _{0,3},$
\end{tabular}
\label{solutions1}
\end{equation}%
where $\nu _{0,\pm }$, $\nu _{0,3}$ are constants. To ensure the fermionic %
commutation relations of $B(t),\,B^\dagger(t)$ they have to obey the relations
$\nu _{0,-}\,\nu _{0,+} + \nu _{0,3}^{2}/4 =0$ \,\, and \,\, $ |\nu
_{0,-}|^{2} + |\nu _{0,+}|^{2} + |\nu _{0,3}|^{2}/2  = 1.$
\medskip

\section{CS for the fermion forced oscillator}

We define coherent states (CS) for a given fermion system as eigenstates of
the corresponding invariant fermion annihilation (or creation) operator $B(t)
$. Since the most general fermion one mode Hamiltonian operator is of the
form of (nonstationary) forced oscillator (\ref{Hf 2}), the one-mode fermion
CS are defined as eigenstates of the invariant ladder operator $B(t)$
(eqs. (\ref{B}), (\ref{constr})):
\begin{equation}
B(t)|\zeta ;t\rangle =\zeta |\zeta ;t\rangle .  \label{|z;t> 1}
\end{equation}%
Since $B(t)$ is invariant operator,  the eigenvalue $\zeta $ does
not depend on time $t$. In terms of the $\zeta $, $B(t)$, $B^{\dagger }(t)$
and the $B(t)$-vacuum $|0;t\rangle $ we have for $|\zeta ;t\rangle $ the
same formulas as for the canonical fermion CS $|\zeta \rangle $ which are
defined {\normalsize \cite{Klauder, Abe89, Maam92, Junker98, Cahill99}} as
\begin{equation}
\left\vert \zeta \right\rangle =e^{-\frac{1}{2}\zeta ^{\ast }\zeta }\left(
\left\vert 0\right\rangle -\text{ }\zeta \left\vert 1\right\rangle \right)
\,.  \label{|z>}
\end{equation}%
where the eigenvalue $\zeta $ is a Grassmannian variable: $\zeta ^{2}=0,\
\zeta \zeta ^{\ast }+\zeta ^{\ast }\zeta =0$, $\left\vert 0\right\rangle $
is the fermionic vacuum, $b\left\vert 0\right\rangle =0$, and $\left\vert
1\right\rangle $ is the one-fermion state, $\left\vert 1\right\rangle
=b^{\dagger }\left\vert 0\right\rangle $. In particular
\begin{equation}
|\zeta ;t\rangle = \re^{-\tfrac{1}{2}\zeta ^{\ast }\zeta }\left( |0;t\rangle
-\zeta B^{\dagger }(t)|0;t\rangle \right) .  \label{|z;t> 2}
\end{equation}%
It remains therefore to construct the (normalized) new ground state $%
|0;t\rangle $ according to its defining equations
\begin{equation}
\begin{aligned} B(t) |0;t\rangle = 0 ,\quad 
\ri\frac{\rd}{dt} |0;t\rangle = H_{f}|0;t\rangle. \end{aligned}  \label{|0;t> 1}
\end{equation}%
We put
\begin{equation}
|0;t\rangle =\alpha _{0}(t)|0\rangle +\alpha _{1}(t)|1\rangle ,
\label{|0;t> 2}
\end{equation}%
substitute this into (\ref{|0;t> 1}) and after some tedious calculations find%
\begin{eqnarray}
\alpha _{1}(t) &=&\alpha _{0}(t)\frac{\nu _{3}^{\ast }(t)}{2\nu _{+}^{\ast
}(t)}, \\
\alpha _{0}(t) &=&\sqrt{|\nu _{+}(t)|}\exp \left[ -\tfrac{\ri}{2}\left(
\varphi _{\nu _{+}}(t)+\tint_{0}^{t}(2g(\tau )+\omega (\tau ))\rd\tau \right) %
\right] ,
\end{eqnarray}%
where $\varphi _{\nu _{+}}$ is the phase of $\nu _{+}(t)$. The state $|\zeta
;t\rangle $ will represent the exact time evolution of an initial canonical
CS $|\zeta \rangle $ if the initial conditions (\ref{nu_i0}) are imposed: %
$|\zeta ;0\rangle = |\zeta \rangle $. In this case, the time evolved state %
$ |\zeta ;t\rangle $ could be again an eigenstate of $b$ if the oscillator %
is not 'forced', i.e. if $f(t)=0$.
Let us note that the time-dependence of the constructed states is obtained %
in terms of solutions to the system of auxiliary equations (\ref{(a)})-(\ref{(c)}),%
or equivalently to the 'classical oscillator' equation (\ref{eps eq}).

Our method of construction of dynamical invariants differs slightly from the
Lewis-Riesenfeld method \cite{Lewis69} (developed for bosonic oscillators).
Lewis and Riesenfeld used to first construct Hermitian invariant, which then
is represented as a product of normally ordered ladder operators. To make
connection to their approach let us suppose that we first succeeded to
construct the Hermitian invariant $N(t)$ and to find some ladder operators $%
\tilde{B}(t)$, $\tilde{B}^{\dagger }(t)$ that factorize it: $N(t)=\tilde{B}%
^{\dagger }(t)\tilde{B}(t)$. It is clear that $\tilde{B}(t)$ may differ from
our non-Hermitian invariant $B(t)$ in a phase factor:\thinspace\ $\tilde{B}%
(t) = {\rm e}^{\ri\varphi (t)}B(t)$.\newline
We can then in a standard way construct normalized eigenstates of $N(t)$,
\begin{equation}
N(t)\widetilde{|0;t\rangle }=0,\quad N(t)\widetilde{%
|1;t\rangle }=\widetilde{|1;t\rangle },  \label{tld 2}
\end{equation}%
and of $\tilde{B}(t)$, \
\begin{equation}
\tilde{B}(t)\widetilde{|\zeta ;t\rangle }=\zeta \widetilde{|\zeta ;t\rangle },
\end{equation}%
\begin{equation}
\widetilde{|\zeta ;t\rangle }=\left( 1-\tfrac{1}{2}\zeta ^{\ast }\zeta
\right) \left[ \widetilde{|0;t\rangle }-\zeta \widetilde{|1;t\rangle }\right]
\label{tld 3}
\end{equation}%
which however do not obey the Schr\"odinger equation since, in general $ \tilde{B}(t)$
may not be invariant. To obtain solutions $|n;t\rangle $ and $ |\zeta ;t\rangle $ the
above eigenstates $\widetilde{|n;t\rangle }$, $n=0,1$, should also be multiplied by
phase factors,
\begin{equation}
|n;t\rangle = {\rm e}^{\ri\phi _{n}(t)}\widetilde{|n;t\rangle },\quad n=0,1,
\label{tld 4}
\end{equation}%
\begin{equation}
|\zeta ;t\rangle =\left( 1-\tfrac{1}{2}\zeta ^{\ast }\zeta \right) \left[
{\rm e}^{\ri\phi _{0}(t)}\widetilde{|0;t\rangle }-\zeta {\rm e}^{\ri\phi _{1}(t)}%
\widetilde{|1;t\rangle }\right]  \label{tld 5}
\end{equation}%
which should obey the equations \
\begin{equation}
\tfrac{\rd}{\rd t}\phi _{n}=\widetilde{\langle n;t|}\ri\tfrac{\partial }{\partial t}%
-H\widetilde{|n;t\rangle }.  \label{tld 6}
\end{equation}%
Evidently the state (\ref{tld 5}) is an eigenstate of $\tilde{B}(t)$ with time
dependent eigenvalue $\zeta (t)=\zeta \exp (\ri\varphi (t))$, $\varphi
(t)=\phi _{1}(t)-\phi _{0}(t)$.\newline
The phase $\varphi (t)=\phi _{1}(t)-\phi _{0}(t)$ consists of two parts -
geometrical one $\varphi ^{G}$, and dynamical one $\varphi ^{D}=\varphi
-\varphi ^{G}$ \cite{Maam99},
\begin{eqnarray}
\varphi ^{G}(t)  &=&\varphi (t)+\int_{0}^{t}\left( \widetilde{\langle 1;t^{\prime }|}H%
\widetilde{|1;t^{\prime }\rangle }-\widetilde{\langle 0;t^{\prime }|}H%
\widetilde{|0;t^{\prime }\rangle }\right) \rd t^{\prime }.
\end{eqnarray}

\subsection*{Concluding Remarks}

\ \ \ In this article, we have studied fermionic system of nonstationary
forced oscillator and we have constructed invariant ladder operators and the
related Fock and coherent states. We succeeded to express these invariants
and the time evolution of the corresponding states in terms of the same
classical equation, that describes the evolution of coherent states of the
boson nonstationary (forced) oscillator \cite{MMT70, Holz70}. The relation of
the invariant ladder operators method to the Lewis-Riesenfeld method \cite{Lewis69}
was briefly described on the example of nonstationary fermion systems. \

\end{document}